\documentclass[11pt,epsf]{article}
\usepackage{hyperref}
\usepackage{subfigure}
\usepackage{graphics}
\usepackage{amssymb}
\usepackage{epsf}
\usepackage{amsmath}
\usepackage{array}
\usepackage{rotating}
\usepackage{pstricks}

\def\a{\alpha}
\def\b{\beta}

\newcommand{\sq}{\square}
\newcommand{\bea}{\begin{eqnarray}}
\newcommand{\eea}{\end{eqnarray}}
\newcommand{\bann}{\begin{eqnarray*}}
\newcommand{\eann}{\end{eqnarray*}}
\newcommand{\bmi}{\begin{minipage}}
\newcommand{\emi}{\end{minipage}}

\newcommand{\beqa}{\begin{eqnarray}}
\newcommand{\eeqa}{\end{eqnarray}}
\def\beq{\begin{equation}}
\def\eeq{\end{equation}}
\def\nn{\nonumber}

\begin{document}
\begin{center}
\vspace{4.cm}
{\bf \large
The Trace Anomaly and the Couplings of QED and QCD to Gravity\footnote{Presented at QCD at work 2010, Beppe Nardulli Memorial Workshop, 20-23 June, Martina Franca, Italy}  }

\vspace{1cm} 

{\bf Roberta Armillis, Claudio Corian\`{o} and  Luigi Delle Rose\footnote{roberta.armillis@le.infn.it, claudio.coriano@le.infn.it, luigi.dellerose@le.infn.it}}

\vspace{1cm}

{\it Dipartimento di Fisica, Universit\`a del Salento \\
and  INFN Sezione di Lecce, Via Arnesano 73100 Lecce, Italy
}\\
\vspace{.5cm}

\begin{abstract}
We report on the computation of the effective actions describing the interaction of gravity both for an abelian and a non-abelian gauge theory, mediated by the trace anomaly. 
\end{abstract}
\end{center}
\newpage

\section{Introduction} 
 In a rather recent work Giannotti and Mottola \cite{Giannotti:2008cv} have pointed out that the effect of the trace anomaly in QED is in the appearance of an anomaly pole in the correlator of the energy momentum tensor $(T)$ with two vector 
 currents $J$, which indicates the existence of additional scalar degrees of freedom in the effective action that describes the coupling of gravity to a gauge theory. Their elaboration goes quite far, by showing that these massless exchanges are already present in a variational solution of the anomaly equation proposed long ago by Riegert \cite{Riegert:1984kt}, solution which is indeed supported in a perturbative framework by an analysis of the corresponding anomaly graphs. 
 
 In the case of anomalous gauge theories a similar pattern emerges, well known since the work of Dolgov and Zakharov 
 \cite{Dolgov:1971ri}, who showed the appearance of similar poles in the spectral density of the AVV gauge anomaly amplitude. More generally, the poles can also be extracted at 1 loop by a decomposition of the anomaly amplitude in terms of longitudinal and transverse form factors \cite{Knecht:2003xy,Jegerlehner:2005fs}, the longitudinal one being responsible for the anomaly and characterized explicitly by a 
 massless pole. An off-shell computation and a mapping from Rosenberg's form of the anomaly graph into the longitudinal/transverse formulation supports these conclusions \cite{Armillis:2009sm}.

This previous analysis and the correspondence with the results of  \cite{Giannotti:2008cv}, extended to the computation of the off shell correlator \cite{Armillis:2009pq} has brought us to conclude \cite{Armillis:2009im} that anomaly poles are the common signature of gauge and conformal anomalies. It does not take a big leap to probably come to similar conclusions also in regard to gravitational anomalies, where again, one may expect the appearance of massless exchanges of similar type, although an explicit computation, in this case, is still missing.  

The perturbative analysis of QED has been recently extended by us to QCD \cite{Armillis:2010qk}, by computing the $TJJ$ correlator 
in a general kinematical domain, which provides more general results respect to the dispersive approach.  
The massless poles found in the study of anomalous gauge theories and in the $TJJ$ correlator are indeed generic contributions, present under a general kinematics, not necessarily linked to the infrared limit of an anomaly amplitude. In fact off-shell correlators are equally characterized by pole contributions also in the UV region \cite{Armillis:2009pq}.
 
\section{The gravitational coupling of gauge theories and the trace anomaly}
Massless poles describe long range interactions, 
 probably accounting for a phase of the effective theory - in this case of a gauge theory coupled to gravity -  which is not yet fully understood at a phenomenological level, probably characterizing some mechanism of condensation. 
 On this point, we just observe that for gauge anomalies, the derivative coupling of the anomaly pole to the anomalous gauge current can be traded with two pseudoscalars of St\"uckelberg type \cite{Armillis:2008bg,Coriano:2008pg} (two gauged axions), one of them ghost-like. The appearance of a ghost in the spectrum is clearly the sign of an instability of the theory, here detected at a perturbative level. We just mention that ghost condensation has received some attention in the past \cite{ArkaniHamed:2003uy}, and some of those ideas, concerning infrared modifications of gravity, may apply to the auxiliary field formulation of these effective actions.  Now we briefly go over a summary of the analysis of the $TJJ$  correlator in QED and QCD before coming to our conclusions.

One well known result of quantum gravity is that the effective action of the trace anomaly is given by a nonlocal form when expressed in terms of the spacetime metric $g_{\mu\nu}$. This was obtained \cite{Riegert:1984kt} from a variational solution of the equation for the trace anomaly \cite{Duff:1977ay}
\bea
T^\mu_\mu =   b \, F + b^{\prime} \, \left( E - \frac{2}{3} \, \square \, R\right) + b'' \, \square \, R +  c\, F^{a \, \mu \nu} F^a_{\mu \nu},
\label{var2}
\eea
(see also \cite{Armillis:2010qk} for more references) which in $D=4$ spacetime dimensions takes the form
\bea
&& \hspace{-.6cm}S_{anom}[g,A] = \label{Tnonl}\\
&&\frac {1}{8}\int d^4x\sqrt{-g}\int d^4x'\sqrt{-g'} \left(E - \frac{2}{3} \square R\right)_x
 \Delta_4^{-1} (x,x')\left[ 2b\,F
 + b' \left(E - \frac{2}{3} \square R\right) + 2\, c\, F_{\mu\nu}F^{\mu\nu}\right]_{x'}. \nonumber
 \label{var1}
\eea
Here, the parameters $b$ and $b'$ are the coefficients of the Weyl tensor squared,
\beq
F = C_{\lambda\mu\nu\rho}C^{\lambda\mu\nu\rho} = R_{\lambda\mu\nu\rho}R^{\lambda\mu\nu\rho}
-2 R_{\mu\nu}R^{\mu\nu}  + \frac{R^2}{3}
\eeq
 and the Euler density
\beq
E = ^*\hskip-.2cmR_{\lambda\mu\nu\rho}\,^*\hskip-.1cm R^{\lambda\mu\nu\rho} =
R_{\lambda\mu\nu\rho}R^{\lambda\mu\nu\rho} - 4R_{\mu\nu}R^{\mu\nu}+ R^2
\eeq
 respectively
of the trace anomaly in a general background curved spacetime.
 
Expanding around flat space, the local formulation of Riegert's action, as shown in \cite{Giannotti:2008cv,Mottola:2006ew}, can be rewritten in the form
\beq
S_{anom}[g,A]  \rightarrow  -\frac{c}{6}\int d^4x\sqrt{-g}\int d^4x'\sqrt{-g'}\, R_x
\, \square^{-1}_{x,x'}\, [F_{\alpha\beta}F^{\alpha\beta}]_{x'}\,,
\label{SSimple}
 \eeq
which is valid to first order in the fluctuation of the metric around a flat background, denoted as $h_{\mu\nu}$
\beq
g_{\mu\nu}= \eta_{\mu\nu} +\kappa h_{\mu\nu}, \qquad\qquad \kappa=\sqrt{16 \pi G_N},
\eeq
with $G_N$ being the 4-dimensional Newton's constant.
 The formulation in terms of auxiliary fields of this action gives  \cite{Giannotti:2008cv}
\beq
S_{anom} [g,A;\varphi,\psi'] =  \int\,d^4x\,\sqrt{-g}
\left[ -\psi'\sq\,\varphi - \frac{R}{3}\, \psi'  + \frac{c}{2} F_{\alpha\beta}F^{\alpha\beta} \varphi\right]\,,
\label{effact}
\eeq
where $\phi$ and $\psi$ are the auxiliary scalar fields. They satisfy the equations
\bea
&&\psi' \equiv  b\, \sq\, \psi\,, \label{diffeq}\\
&&\square\,\psi' =  \frac{c}{2}\, F_{\alpha\beta}F^{\alpha\beta} \,,\\
&&\square\, \varphi = -\frac{R}{3}\,.
\eea

\begin{figure}[t]
\begin{center}
\includegraphics[scale=0.9]{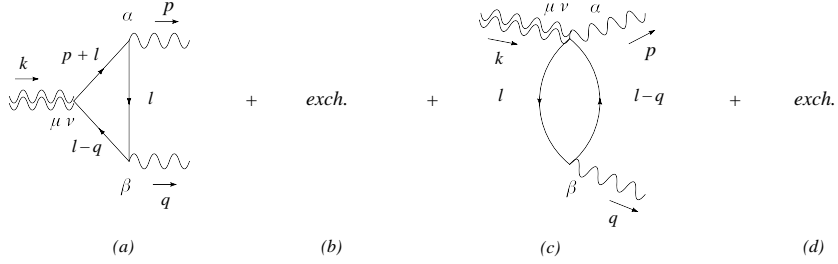}
\caption{\small The fermionic contributions  with a graviton $h_{\mu\nu}$ in the initial state and two gluons $A^a_\a, A^b_\b$ in the final state. }
\label{fermloop}
\end{center}
\end{figure}
The $TJJ$ amplitude is shown in Fig. \ref{fermloop} in the QED case, with additional contributions that appear in the QCD case. 
The off shell tensor analysis of these diagrams is rather cumbersome, but the amplitude can be arranged in terms of 13 invariant amplitudes 
\bea
\Gamma^{\mu\nu\alpha\beta}(p,q) =  \, \sum_{i=1}^{13} F_i (s; s_1, s_2,m^2)\ t_i^{\mu\nu\alpha\beta}(p,q)\,,
\label{Gamt}
\eea
where the form factors $F_i$ are functions of the kinematical invariants $s=k^2=(p+q)^2$, $s_1=p^2$, $s_2=q^2$ and of the internal mass $m$. Explicit expressions of these form factors are given in \cite{Armillis:2009pq}. In the massless case only few form factors survive and one gets
\bea
F_{1} (s, 0, 0, 0) &=& - \frac{e^2}{18 \pi^2  s}, \\
F_{3} (s, 0, 0, 0) &=&  F_{5} (s, 0, 0, 0) = - \frac{e^2}{144 \pi^2 \, s}, \\
F_{7} (s, 0, 0, 0) &=& -4 \, F_{3} (s, 0, 0, 0), \\
F_{13, R} (s, 0, 0, 0) &=& - \frac{e^2}{144 \pi^2} \, \left[ 12 \log \left(-\frac{s}{\mu^2}\right) - 35\right],
\eea
where $F_{13\, R}$ denotes the renormalized amplitude. The anomaly is entirely given by $F_1$, which indeed shows the presence of an anomaly pole. Further details on the organization of the effective action mediated by the trace anomaly can be found 
in  \cite{Armillis:2009pq}.

Coming to QCD, the full on-shell vertex, which is the sum of the quark and pure gauge contributions, can be decomposed by using three appropriate tensor structures $\phi_i^{\mu\nu\alpha\beta}$, given in \cite{Armillis:2010qk}, and appearing in the expansion of quark ($\Gamma_q^{\mu\nu\a\b}(p,q)$) and gluon ($\Gamma_g^{\mu\nu\a\b}(p,q)$) subsets of diagrams
\bea
\Gamma^{\mu\nu\alpha\beta}(p,q) =  \Gamma^{\mu\nu\alpha\beta}_g(p,q) + \Gamma^{\mu\nu\alpha\beta}_q(p,q) =  \sum_{i=1}^{3} \Phi_{i} (s,0, 0)\, \delta^{ab}\, \phi_i^{\mu\nu\alpha\beta}(p,q)\,,
\eea
with form factors defined as
\bea
\Phi_i(s,0,0) = \Phi_{i,\,g}(s,0,0) + \sum_{j=1}^{n_f}\Phi_{i, \,q}(s,0,0,m_j^2),
\eea
where the sum runs over the $n_f$ quark flavors. In particular we find
\bea
\Phi_{1}(s,0,0) &=& - \frac{g^2}{72 \pi^2 \,s}\left(2 n_f - 11 C_A\right) + \frac{g^2}{6 \pi^2}\sum_{i=1}^{n_f} m_i^2 \, \bigg\{ \frac{1}{s^2} \, - \, \frac{1} {2 s}\mathcal C_0 (s, 0, 0, m_i^2)
\bigg[1-\frac{4 m_i^2}{ s}\bigg] \bigg\}, \,
\label{Phi1}\\
\Phi_{2}(s,0,0) &=& - \frac{g^2}{288 \pi^2 \,s}\left(n_f - C_A\right) \nn \\
&&- \frac{g^2}{24 \pi^2} \sum_{i=1}^{n_f} m_i^2 \, \bigg\{ \frac{1}{s^2}
+ \frac{ 3}{ s^2} \mathcal D (s, 0, 0, m_i^2)
+ \frac{ 1}{s } \mathcal C_0(s, 0, 0, m_i^2 )\, \left[ 1 + \frac{2 m_i^2}{s}\right]\bigg\},
\label{Phi2} \\
\Phi_{3}(s,0,0) &=& \frac{g^2}{288 \pi^2}\left(11 n_f - 65 C_A\right) - \frac{g^2 \, C_A}{8 \pi^2} \bigg[ \frac{11}{6} \mathcal B_0^{\overline{MS}}(s,0) - \mathcal B_0^{\overline{MS}}(0,0) +  s  \,\mathcal C_0(s,0,0,0) \bigg] \nn \\
&& + \, \frac{g^2}{8 \pi^2} \sum_{i=1}^{n_f}\bigg\{  \frac{1}{3}\mathcal B_0^{\overline{MS}}(s, m_i^2) + m_i^2 \, \bigg[
\frac{1}{s}
 + \frac{5}{3 s}  \mathcal D (s, 0, 0, m_i^2) + \mathcal C_0 (s, 0,0,m_i^2) \,\left[1 + \frac{2 m_i^2}{s}\right]
\bigg]\bigg\} ,\nn \\
\label{Phi3}
\eea
with $C_A = N_C$. The scalar integrals $\mathcal B_0^{\overline{MS}}$, $\mathcal D$ and $\mathcal C_0$ are defined in \cite{Armillis:2010qk}.
Notice the appearance in the total amplitude of the $1/s$ pole in $\Phi_1$, which is present both in the quark and in the gluon sectors, and which saturates the contribution to the trace anomaly in the massless limit. In this case the entire trace anomaly is just proportional to this component, which becomes 
\beq
\Phi_{1}(s,0,0) = - \frac{g^2}{72 \pi^2 \,s}\left(2 n_f - 11 C_A\right).
\label{polepole}
\eeq
Further elaborations show that the effective action is given by
\bea
S_{pole} &=&	- \frac{c}{6}\, \int d^4 x \, d^4 y \,R^{(1)}(x)\, \square^{-1}(x,y) \,  F^a_{\alpha \beta} \,  F^{a \, \alpha \beta} \nn\\
&=& \frac{1}{3} \, \frac{g^3}{16 \pi^2} \left (  - \frac{11}{3} \, C_A + \frac{2}{3} \, n_f \right)  \, \int d^4 x \, d^4 y \,R^{(1)}(x)\, \square^{-1}(x,y) \, F_{\alpha \beta}F^{\alpha \beta}
\eea
and is in agreement with Eq. (\ref{SSimple}), derived from the nonlocal gravitational action (\ref{var1}).
Here $R^{(1)}$ denotes the linearized expression of the Ricci scalar 
\beq
 R^{(1)}_x\equiv \partial^x_\mu\, \partial^x_\nu \, h^{\mu\nu} - \square \,  h, \qquad h=\eta_{\mu\nu} \, h^{\mu\nu}
 \eeq
and the constant $c$ is related to the non-abelian $\beta$ function as
\beq
c= - 2 \, \, \frac{\beta (g)}{g}.
\eeq
\section{Conclusions}
The possible significance of the effective degrees of freedom described in these actions is still open, with suggestions 
that touch upon the role of QCD in solving the dark energy problem of cosmology \cite{Urban:2009vy, Urban:2009yg}. 
In fact, it has been suggested that the small value for the vacuum energy density originally attributed to the anomaly 
\cite{Starobinsky:1980te}, could be raised to the expected one $(10^{-3} \textrm{eV})^4$ if the gravitational effective action is characterized by some effective nonlocality. In this case the contribution due to the trace anomaly could be modified \cite{Klinkhamer:2009nn}.  Other possible extensions of this line of research concerns the case of anomaly mediation in supersymmetric theories (see for instance \cite{Jung:2009dg}).

\centerline{Acknowledgments} 
This work is supported in part  by the European Union through the Marie Curie Research and Training Network Universenet (MRTN-CT-2006-035863).

\end{document}